\documentclass[doublecol]{epl2} 
\usepackage{subcaption}
\title{Stability of electrodynamically levitated one or many charged droplet in presence of noise}
\author{Mohit Singh\inst{1} \and Neha Gawande\inst{1} \and Y.S. Mayya\inst{1} \and Rochish Thaokar\inst{1}}
\shortauthor{M. Singh \etal}

\institute{                    
	\inst{1} Department of Chemical Engineering, Indian Institute of Technology Bombay, Mumbai, India.\\
	}

\date{\today}

\abstract{
The theory of the effect of external fluctuation force on the stability and spatial distribution of mutually interacting and slowly evaporating charged drops, levitated in an electrodynamic balance, is presented using classical pseudo-potential approach. The theory is supplemented with numerical simulations where the non-homogeneous modified Mathieu equation is solved for single droplet as well as many droplets. The transition from the well ordered Coulombic crystal to randomly distributed liquid like structure is observed above a threshold value of the order parameter. The theory and simulations are found to be in fair agreement with each other. The simulation is aimed at studying the stability of structures for capturing the pollutant particle form the air streams using contactless membrane. (The manuscript is submitted in EPL (Europhysics Letters) journal)}  
\begin{document}
\maketitle
\section{Introduction}
The understanding and the limitations of dynamically trapping of subatomic charged microparticles in the Penning traps\cite{wineland1973, brown1986} and ions in the Paul trap \cite{dehmelt1968, paul1990} in vacuum have been a subject of detailed studies in the past. However, comparatively, fewer\cite{winter1991, wuerker1959, singh2017} works are available on the levitation of charged micron and sub-micron sized liquid drops in an electrodynamic levitator at atmospheric conditions near standard temperature and pressure (STP). While the noise is expected to play a significant role in the dynamics of ions and elementary particles in electrodynamic traps, most studies (\cite{winter1991, wuerker1959, singh2017} barring few \cite{arnold1993}) have ignored the effect of thermal noise on the levitation dynamics of micron and submicron sized liquid droplets.

Towards this Arnold s. \etal \cite{arnold1995, arnold1993} considered a charged micron sized drop in a quadrupole Paul trap near STP and introduced the effect of thermal noise in the resulting modified Mathieu equation which was akin to a Langevin equation. An equivalent Fokker-Planck equations was realized and solved numerically. A good agreement was reported between numerical results and their experimental data. For experimental observations, they trapped micrometre-sized drops in a Paul trap ($2r_0$=9 mm) in the presence of an AC electric bias voltage $V$=1.0$V$ and at 60 $Hz$ applied frequency which resulted in $\leq$ 1.0 $\mu$m fluctuations. Similarly, Blatt R. \etal\cite{blatt1986} and Zerbe C. \etal\cite{zerbe1994} computed the thermal fluctuations in the position and velocity of the drop by solving the Fokker-Planck equation in the presence of a gas medium in the limit of small magnitude of parametric modulation parameter. Joos E. \etal \cite{joos1989} solved the Langevin equation in the limit of small $q$ parameter where $q$ was potential parameter and derived the series expansions of the thermal fluctuations of position and velocity. The understanding of thermal noise on the dynamics of a single levitated droplet is the first step in investigating its effect and importance in many droplet system.

The levitation of multiple droplets in a trap \cite{wuerker1959, aardahl1997, singh2017} can lead to the formation of beautiful colloidal crystals wherein the addition of Coulombic interaction can lead to the formation of well ordered arrays of droplets, often exhibiting degeneracy. It is then interesting to investigate the effect of noise on such ordered structures. Another motivation for examining the effect of noise on droplets pertains to uncontrolled interactions in many droplet levitation systems. It is found that many of the levitated droplets are relatively stable forming a membrane like 2D structure. There are possible technological applications such as non-contact membrane development which has potential applications, for example, in air cleaning and cloud formation studies. However, in many body systems, small positional fluctuations could lead to random fluctuations in the electric field of a levitated droplet. Thus because of the large degrees of the freedom in such a many body system, it is then interesting to know how in addition to electrical fluctuations the perturbation induced by convection flows or some other ways such as the presence of other ions could effectively affect the stability of the overall many body system. This motivates us to examine the effect of noise not only for low intensity (i.e. thermal fluctuations) but also for large noise (mechanical or athermal) amplitudes. In this context, another question arises, whether such fluctuations can induce changes in the structure arrangement such as transition form well organized to a randomly distributed structures akin to thermodynamic \textquotedblleft phase transition\textquotedblright. Keeping these objectives in mind we performed the analysis of the effect of noise not only on the single droplet system but also in many droplets which seems to indicate the possibility of phase change behaviour. We employed both numerical simulations and analytical theory to address the following relevant questions
\begin{itemize}
	\item What will be the functional relationship between the strength of the noise and the mean variance in case of a single droplet as well as two droplet system? 
	\item What will be the stability limit of the single droplet as well as two droplets in the presence of noise?
	\item What is the threshold limit of the strength of the noise for transition from well organized to random state? 
	\item What will be the structural arrangement in case of a large number ($\sim$100) of drops in the presence of noise? 
\end{itemize}
Although the theoretical analysis of fluctuations in the present manuscript is done for a thermal system it can be mapped to the external or athermal noise as long as the Gaussian white noise characteristics are maintained. Unlike Arnol S. \etal \cite{arnold1985} we have theoretically investigated the stability of single droplet as well as two droplet system in the presence of noise by relating the potential energy obtained from the classical pseudo approach with the Boltzmann probability distribution by involving an effective temperature. The mean variance as a function of strength noise is then obtained by comparing the expression of probability distribution with the standard Gaussian distribution function. Since the theoretical development lies on the separation of total motion into slowly varying and fast oscillatory components of motion it is difficult to identify the slow and fast components of motion in the presence of noise. Hence we have employed a simple approach for theoretical development of the process. We determine the effective potential in the deterministic system and obtain the appropriate expression for the variance through an effective temperature. Although the theory is developed up to two droplet system a similar approach can be extended for many body system as well.

\section{Governing equations for charged drops in quadrupole field with noise}
The equation of motion for the levitation of identically charged ($q$) drops and having the same mass ($m$), in a quadrupole balance are discussed in reference \cite{singh2017}. Briefly, the electrodynamics levitator consists of two end cap electrodes and a ring electrode. The end cap electrodes are maintained at the same potential, given by $\phi_0cos(\omega t)$, such that its spatial distribution within the quadrupole set-up is given by the standard form:

\begin{equation}
\phi(r,z,t)=\frac{2z^2-r^2+r{_0}^2}{2z{_0}^2+r{_0}^2}\phi_0cos(\omega t) \label{potential}
\end{equation} 

Where, $r$ is the in plane radial direction, $r^2=x^2+y^2$, $z$ is the vertical direction in a cylindrical coordinate system, $r_0$ is the radius of the ring electrode, $z_0$ is the distance between the centre of a ring electrode and end-cap electrode, $\phi_0$ is the applied potential, $\omega$= 2$\pi$$f$, $f$ is the applied frequency. Consider a drop having charge $q$ and mass $m$ levitated in an alternate quadrupole electric field. The corresponding electric field in the radial ($r$) and vertical ($z$) directions are as follows

\begin{equation}
E_r=-\frac{d\phi}{dr}=\frac{2r}{2z{_0}^2+r{_0}^2}\phi_0cos(\omega t) \label{eler}
\end{equation}

\begin{equation}
E_z=-\frac{d\phi}{dz}=-\frac{4z}{2z{_0}^2+r{_0}^2}\phi_0cos(\omega t) \label{elez}
\end{equation}
In addition to the single drop system the $N$ drop system experiences the force of mutual electrostatic repulsion along with the quadrupole force and the viscous drag force. The equations of motion for a $N$ drop system without fluctuation can be modified by adding a term for accounting the fluctuations such that the resulting equations can be written as,

\begin{equation}
\frac{d^2x_i}{d\tau^2} +  c_i \frac{dx_i}{d\tau} -a_i \thinspace cos(\tau)x_i-\sum_{i\neq j} b_{ij}\frac{x_i-x_j}{\mid r_{ij} \mid^3}=f(\tau)
\label{newtonx}
\end{equation}

\begin{equation}
\frac{d^2y_i}{d\tau^2} +  c_i \frac{dy_i}{d{\tau}} -a_i \thinspace cos(\tau)y_i-\sum_{i\neq j} b_{ij}\frac{y_i-y_j}{\mid r_{ij} \mid^3}=f(\tau)
\label{newtony}
\end{equation}

\begin{equation}
\frac{d^2z_i}{d\tau^2} +  c_i \frac{dz_i}{d{\tau}} +2a_i \thinspace cos(\tau)z_i-\sum_{i\neq j} b_{ij}\frac{z_i-z_j}{\mid r_{ij} \mid^3}=f(\tau)
\label{newtonz}
\end{equation}

where, $\tau$=$\omega t$, $a_i(=\frac{2q_i\phi_0}{(2z{_0}^2+r{_0}^2)m_i \omega^2})$ is the stability parameter, $c_i(=\frac{3\eta \pi {D_p}_i}{m_i\omega})$ is the drag coefficient, $b_{ij}=\frac{q_iq_j}{16\pi\epsilon\epsilon_0 m_i \omega^2}$,
$$r_{ij}=\sqrt{(x_i-x_j)^2+(y_i-y_j)^2+(z_i-z_j)^2},$$ $q_i/m_i$ is the charge to mass ration of the $i^{th}$ droplet, $\eta$ is the viscosity of air, $D_{p_i}$ is the diameter of the $i^{th}$ drop,    
and the subscripts $i$ and $j$ represent the $i^{th}$ and $j^{th}$ drop in a $N$ drop system. The term $f(\tau)$ is the non-dimensional form of $f(t)$, which is a random force and will be assumed to be Gaussian zero-mean white noise. The noise is assumed to be isotropic. The properties of such a distribution are determined by the mean and the covariance of $f(t)$ which are $\textless f(t) \textgreater=0$,  $\textless f(t)f'(t) \textgreater=L{_0}'\delta(t-t')$, $f(\tau)$ is the suitable non-dimensional force, $L{_0}=\frac{L{_0}'}{m \omega^3}$. In case of a thermal noise the expression of $L{_0}$ is given by, $\sqrt{\frac{2K_BT}{hm\omega^2}}$, where the $h$ is the time step.          

The dynamical equation of motion for a single drop in an oscillating electric field is described by the Mathieu equation, modified to take into account the viscous drag and fluctuations forces. The other forces such as gravitational and dielectrophoretic forces are neglected by assuming that the drop is very small and the dielectric forces are $\sim$O($a^3$). The force balance equation in the radial direction for a single drop can be given as, 

\begin{equation}
\frac{d^2r}{d\tau^2} +  c \frac{dr}{d{\tau}} -a \thinspace cos(\tau)r=f(\tau)
\label{snewtonr}
\end{equation} 

Where $\tau(=\omega t$) is the dimensionless time, $\omega$= angular frequency of the AC field, $t$ is the time,  $a$ is the trapping parameter, defined as $a=\frac{2q\phi_0}{(2z{_0}^2+r{_0}^2)m \omega^2}$, $c$ is the coefficient of drag (dimensionless), defined as $c=\frac{3\eta \pi D_p}{m\omega}$, $q/m$ is the charge to mass ratio of the drop, $\phi_0$ is the applied electric potential and $D_p$ is the diameter of the drop. 

\section{Single drop system}
The brief over view of basic analytical formulation and corresponding solution of equation of motion for case of $f(\tau) \neq 0 $ \cite{arnold1995, joos1989, blatt1986} \& $f(\tau) = 0 $ \cite{wuerker1959, singh2017} already available in the literature is re-emphasized here. Since the single drop system is not conservatively confined but ponderomotively confined, the analytical solution of equation of motion of a charged drop in the absence of noise can be obtained by the traditional Dehmelt approach (also known as adiabatic or pseudo potential approach) with appropriate assumptions. In this approach, the total motion of the droplet is decomposed into fast $s(\tau)$ and slow $\bar{r}(\tau)$ components. The solution of slowly varying component ($\bar{r}(\tau)$) of the motion depends upon the solution of fast varying component $s(\tau)$. The solution of $s(\tau)$ can be obtained by invoking simple harmonic solution which is $s(\tau)= A \cos(\tau)+B \sin (\tau)$. The constants $A$ \& $B$ can be obtained by comparing the coefficients of $\cos(\tau)$ \& $\sin(\tau)$. Upon inserting the average steady state solution of $s(\tau)$ into the $\bar{r}$ equation results in the following equation of motion, 
\begin{equation}
\ddot{\bar{r}}(\tau)+c\dot{\bar{r}}(\tau)+\frac{1}{2}\frac{a^2\bar{r}}{1+c^2}=0 \label{rbar1}
\end{equation}
The detailed derivation of intermediate steps and assumptions are given in the supplementary file. The last term in eq.~\ref{rbar1} is a time averaged conservative force. The expression for the ponderomotive force at steady state can be written as $f=\frac{1}{2}\frac{a^2 \bar{r}}{1+c^2}$ and the corresponding potential energy is $$E(r)=\int F dr=\frac{1}{2}\frac{a^2 \bar{r}^2}{1+c^2}$$. The dimensional form of the potential energy can be written as,
$$E(r)=\int F dr=\frac{1}{2}\frac{a^2 \bar{r}^2 m \omega^2}{1+c^2}$$
Assuming a Boltzman distribution for the probability distribution for this potential energy, 
$$P(r)=EXP[\frac{-E(r)}{K_BT}]$$ 
For a thermal system, in the case of Gaussian distribution, the distribution is characterized by its standard form, which can be expressed as $P(r)=exp[\frac{-r^2}{2 \sigma{_r}^2}]$, where $P(r)$ is the probability distribution and $\sigma{_r}$ is the variance of the distribution. The total force acting on the drop at steady state in absence of noise will be modulated in case of additional force term in the form of fluctuations. The strength of noise in the system is introduced in the form of Gaussian white noise. Hence the probability distribution in terms of the potential energy can be obtained by simple substitution of the value of $E(r)$ in the equation of probability distribution. The variance of the distribution can be obtained by comparing the expression of probability distribution obtained from the Boltzmann distribution and the standard form of Gaussian distribution ( as given above). The variance can be obtained as, 
\begin{equation}
\sigma{_r}^2=\frac{2K_BT(1+c^2)}{a^2 m \omega^2}   \label{sig}
\end{equation}
Eq.~\ref{sig} is transformed to system dependent parameters, using the definition of the strength of fluctuations i.e $L_0$=$\sqrt{\frac{2 K_BTc}{h m \omega^2}}$, as given in the previous section. Upon substituting the expression of $L_0$ in eq.~\ref{sig}, the variance can be rearranged to express in the form of a equation given below,
\begin{equation}
\sigma{_r}^2=[\frac{L_0}{a}]^2 \frac{h}{a} (1+c^2)  \label{sig1}
\end{equation}
\begin{figure}[htb]
	\centering
	\onefigure[width=0.4\textwidth]{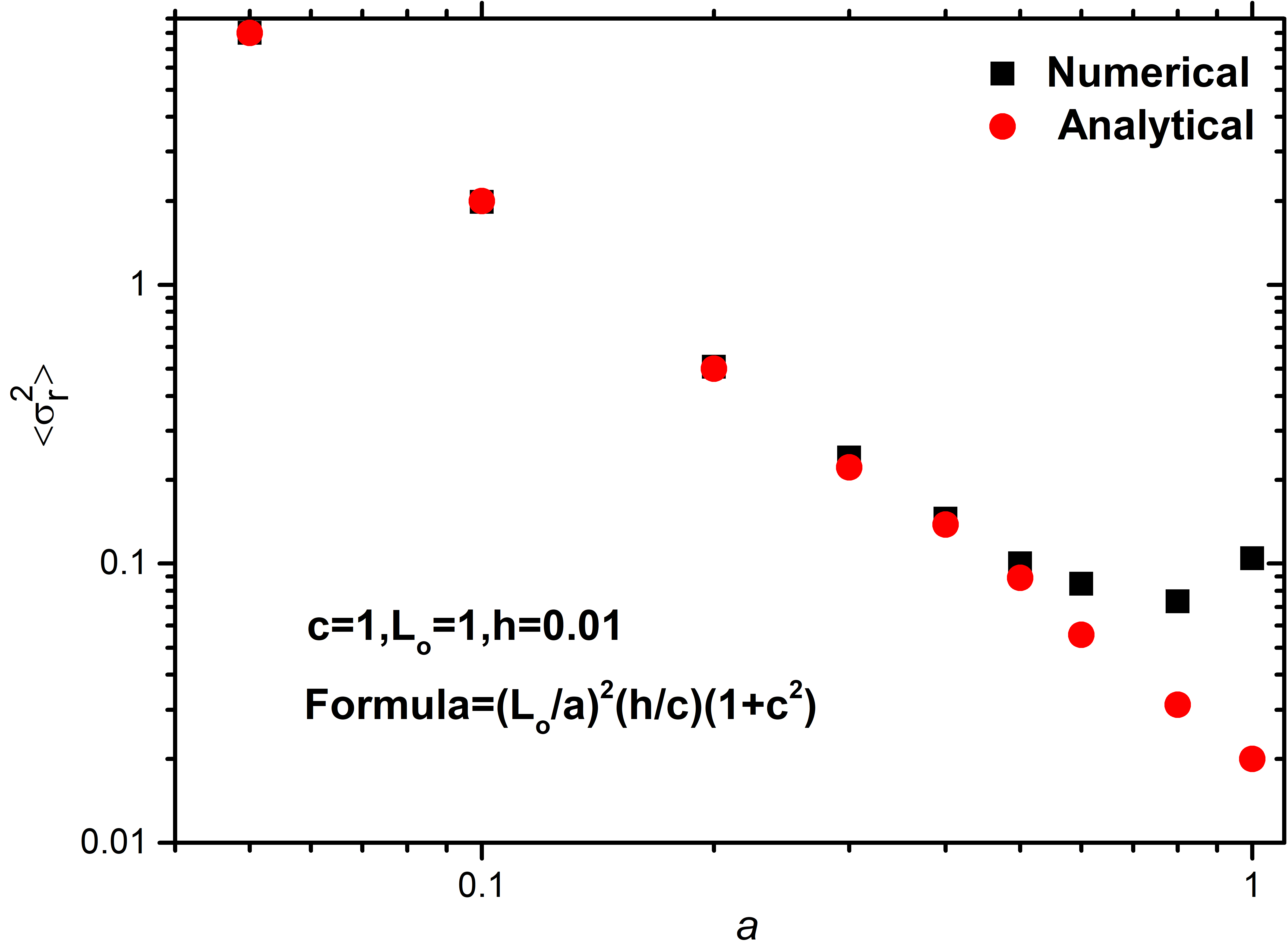}
	\caption{Change in the mean variance of the motion of the single drop in an electrodynamic balance. The typical parameters used for the simulations are given in the inset of the figure}
	\label{fig:progeny2}
\end{figure}

It can be observed from eq.~\ref{sig1} that the $\sigma{_r}^2$ is proportional to $L_0$ and inversely proportional to $a$. Further, to validate the analytically obtained expression of $\sigma{_r}^2$ numerical simulations are performed. The details of the numerical scheme and implementation are provided in Singh M. \etal \cite{singh2017}. The $\sigma{_r}^2$ from the numerical simulations are obtained by taking time average over 1000 trajectories. A single trajectory is obtained by solving the Newtonian dynamics for 2 hundred thousand iterations with 0.01 time step ($h$). The quantitative comparison of the results obtained from the numerical simulations and analytical expression is shown in fig.~\ref{fig:progeny2}. It is observed that the theory and simulation holds a good agreement up to $a$$\sim$$0.5$. This is due to the fact that the higher order terms in simple harmonic solution are neglected. A similar result was reported by Arnold s. \etal \cite{arnold1995} in his fig.~2. Although the theory of a single drop stochastic calculation is well developed in the literature we have used a simpler approach and a fair agreement is observed between theory and simulations. 

In most practical cases one encounters multiple droplets. The complicated interactions in many body system can be best understood in the case of two droplet system. The additional charge interaction term ($b$) makes the equation non-linear and the corresponding analytical solution is expected to be more informative. It is discussed in the literature \cite{singh2017} that the presence of drop-drop interactions does not change the stability diagram. Hence the change in the mean variance of inter droplet separation with the strength of noise will be qualitatively the same. It will be seen that it is indeed the case.  

\section{Two drop system}
Unlike the case of a single droplet, a two drop system admits mutual electrostatic repulsion in addition to the externally applied time varying alternate quadrupole force and the viscous drag force. After a sufficiently long time, the effects of initial conditions such as initial position and initial velocities are expected to diminished such that the droplets undergo stationary radial oscillations. It has been reported \cite{singh2017, singh2018theoretical} that even if one starts with arbitrary initial orientations (in any of the X, Y, Z directions) of the two drops eventually both the droplet would settle down to stationary oscillations in the X-Y plane. Further, in the view of the inter-drop distance, both the droplets oscillate symmetrically (i.e equidistant from the centre) and the line of symmetry passes through the centre of the ring.

Recently, a rigorous theoretical formulation of inter-drop distance in case of two or more than two drop system was reported by Singh M. \etal \cite{singh2018theoretical}. To the best of our knowledge, the theoretical study of the two and more droplet levitation in the presence of external random force is not yet reported and is one of the aims of this work. To explore the effect of noise the basic analytical formulation and corresponding solution of the equation of motion are presented here. Consider the dynamics of two identical drops of mass $m$, charged to identical levels $q$, in an oscillating quadrupole field. The applied potential is $\phi_0$$cos(\tau)$ at the endcap and zero on the ring such that its spatial distribution within the quadrupole set-up is given eq.~\ref{potential}. Then in terms of the inter-drop distance $2r$, the equation of motion for $r$, expressed in terms of suitably non-dimensionalized time and system parameters, can be given by eq.~\ref{newtonx}, \ref{newtony}, \ref{newtonz}. Now substituting $r=|r_1-r_2|$ in eq.~\ref{newtonx} and \ref{newtony}, the equation of motion for $r$ may written as
\begin{equation}
\frac{d^2r_i}{d\tau^2} +  c_i \frac{dr_i}{d\tau} -a_i \thinspace cos(\tau)r_i-\sum_{i\neq j} b_{ij}\frac{r_i-r_j}{\mid r_{ij} \mid^3}=f(t)
\label{newton2r}
\end{equation}
Here $i$=1 and 2 for drop $1^{st}$ \& $2^{nd}$. Expressing eq.~\ref{newton2r} in more general terms in order to avoid notational complications and switch off the fluctuation force by substituting $f(\tau)=0$, the eq.~\ref{newton2r} may rewritten as
\begin{equation}
\ddot{\bar{r}}(\tau)+c\dot{\bar{r}}(\tau)+a r(\tau) cos(\tau)-\frac{b_2}{r(\tau)^2}=0 \label{rbar2d}
\end{equation} 

\begin{figure}
	\centering		
	\begin{subfigure}[b]{0.67\linewidth}
		\onefigure[width=\linewidth]{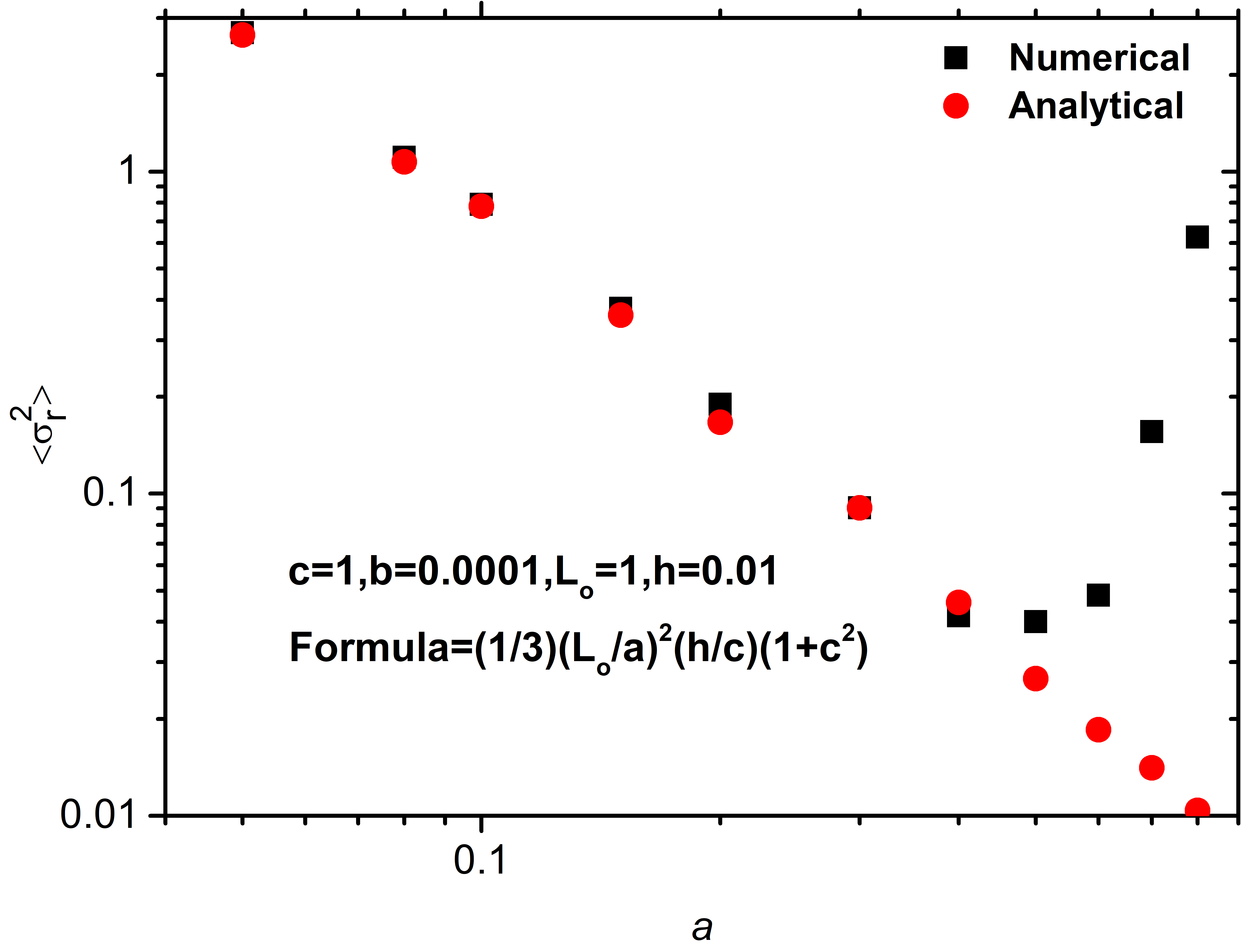}
		\caption{}
		\label{fig:sig2_vs_a}
	\end{subfigure}
   \begin{subfigure}[b]{0.7\linewidth}
		\onefigure[width=\linewidth]{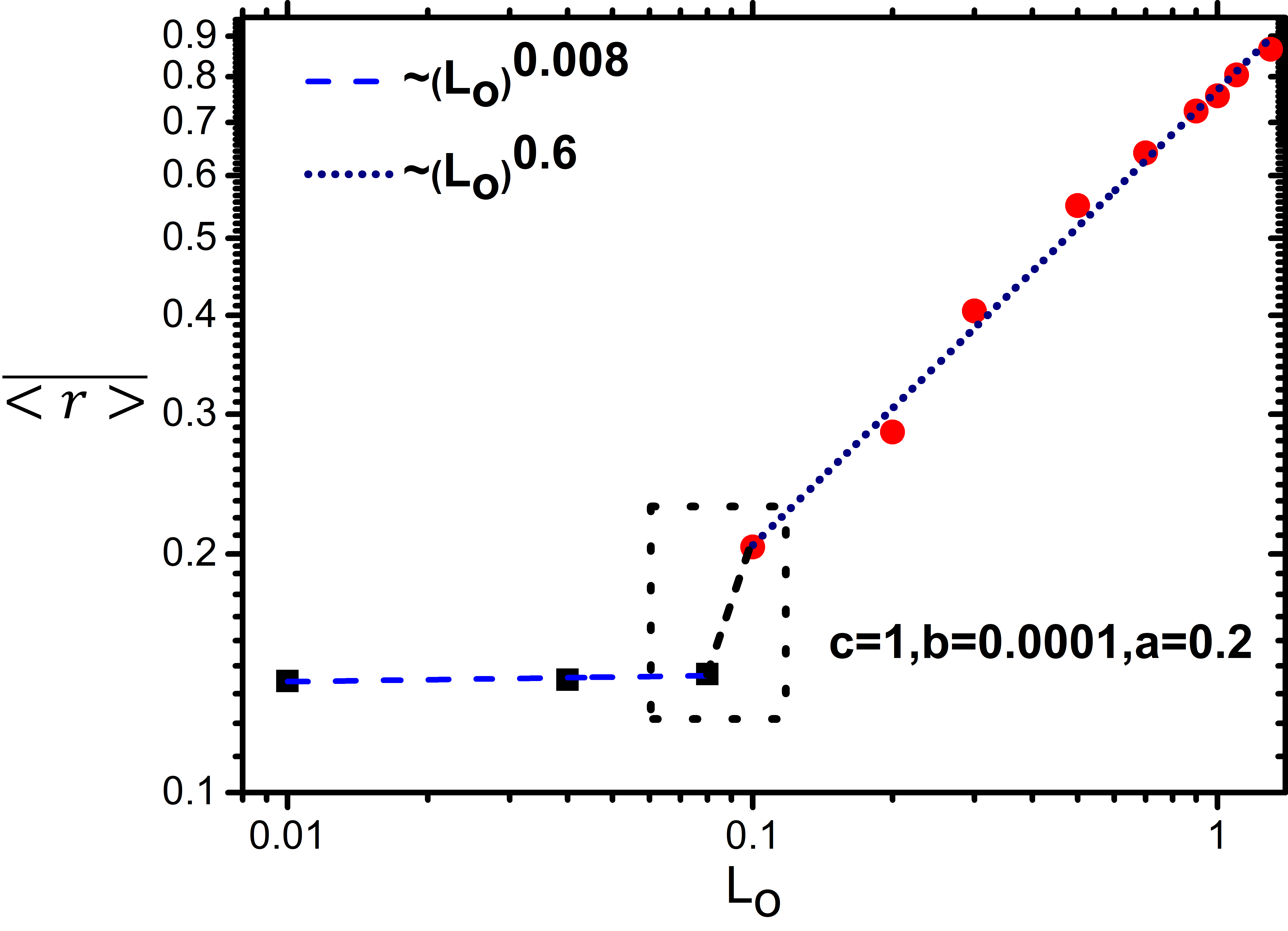}
		\caption{}
		\label{fig:rbar_vs_Lo}
	\end{subfigure}
	\begin{subfigure}[b]{0.7\linewidth}
		\centering
		\onefigure[width=\linewidth]{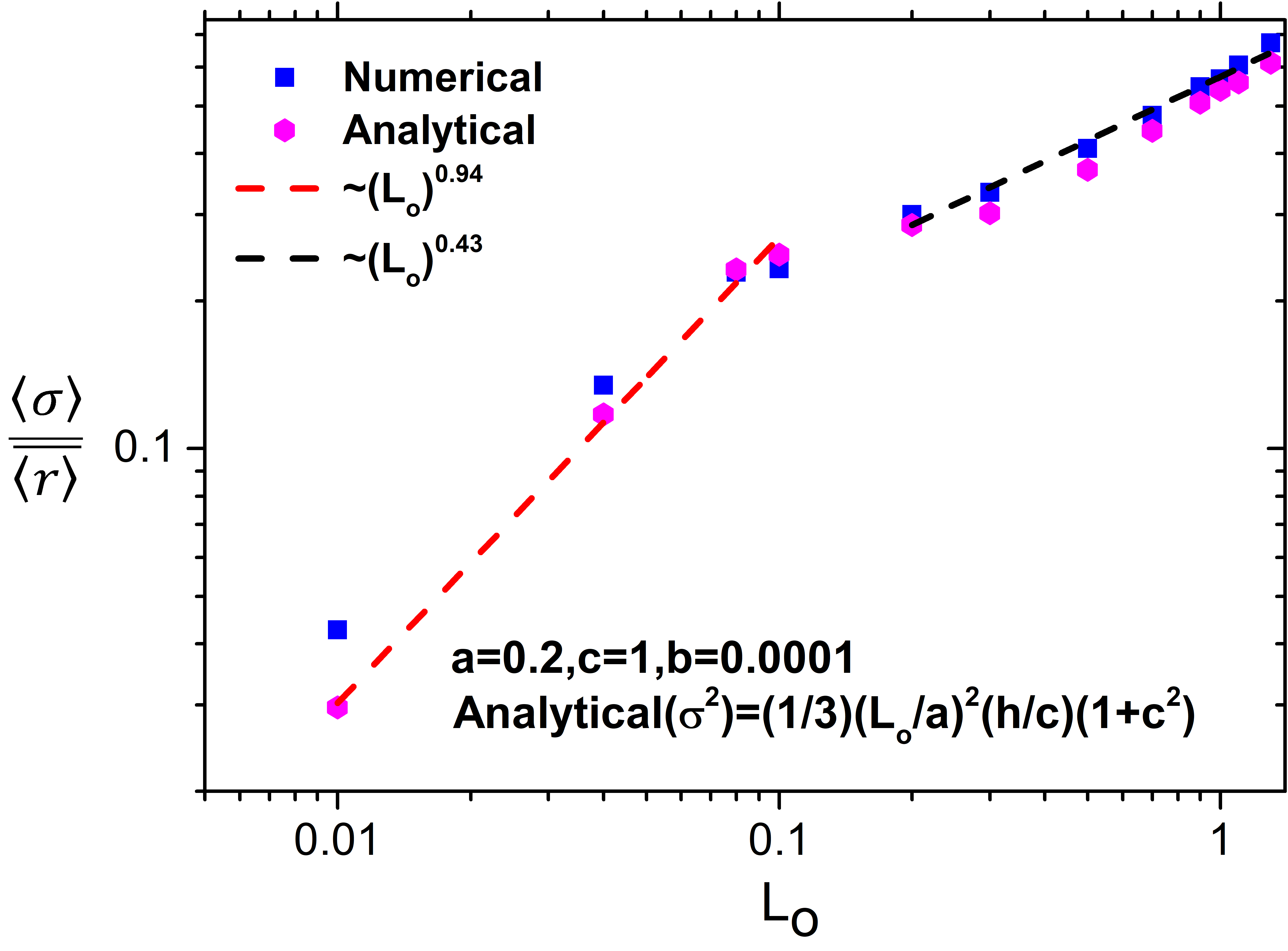}
		\caption{}
		\label{fig:sig_r_vs_lo}
	\end{subfigure}
	\caption{(a) Numerical simulation of change in the mean variance of one of the two drop in an electrodynamic balance, (b) Numerical simulations of the change in the mean inter-drop separation with the strength of the noise, (c) change in $\frac{\langle\sigma\rangle}{\bar{\langle r \rangle}}$ with the noise and comparison with theory} \label{fig:two_n}
\end{figure}

In the above equation (eq.~(\ref{rbar2d})) the parameter $c$ is the drag coefficient and $a$ is the trap parameter for the two drop case, similar to the single drop case. The additional term $b_2$=$l{_c}^3$=$\frac{q^3}{16 \pi \epsilon\epsilon_0 m \omega^2}$ is the fundamental coulomb interaction parameter that arises from the two charged drops interaction or repulsion, having the dimensions of volume. The parameter $l_c$ is the governing length scale for the problem. For a pair of electrostatically repelling drops, $l_c$ provides a more appropriate length scale than the radius of the ring. The average equation of slowly varying component ($\bar{r}$) in case of two droplet system obtained using the Dehmelt approximation as,   
\begin{equation}
\ddot{\bar{r}}(\tau)+c\dot{\bar{r}}(\tau)+\frac{1}{2}\frac{a\bar{r}}{c^2+1}-\frac{b_2}{r(\tau)^2}=0 \label{rbar2d1}
\end{equation}
The detailed derivation of eq.~\ref{rbar2d1} is given in the supplementary file. The above equation of motion is homogeneous i.e. there is no net force acting on the body. Now if one introduces noise in the system in the form of Gaussian white noise ($L{_0}'\delta(\tau-\tau')$). The eq.~\ref{rbar2d1} can be written as  
\begin{equation}
\ddot{\bar{r}}(\tau)+c\dot{\bar{r}}(\tau)+\frac{1}{2}\frac{a\bar{r}}{c^2+1}-\frac{b_2}{r(\tau)^2}=f(\tau) \label{rbar2dnoise}
\end{equation}
The steady state solution of eq.~\ref{rbar2dnoise} is 
\begin{equation}
F_{ND}=\frac{1}{2}\frac{a\bar{r}}{c^2+1}-\frac{b_2}{r(\tau)^2}=f(\tau) \label{rbar2dnoi}
\end{equation}
Where, $F_{ND}$ is the steady state non-dimensional force. The potential energy ($E(r)$) of the corresponding force is 
\begin{equation}
E(r)=\int F_{ND} dr=\frac{1}{4}\frac{a^2\bar{r}^2}{c^2+1}+\frac{b_2}{r(\tau)}
\end{equation}
$$E'(r)=\frac{1}{2}\frac{a^2\bar{r}}{c^2+1}-\frac{b_2}{r(\tau)^2}$$
Hence the minimum or maximum value of potential energy can be obtained by setting the first derivative of potential energy equal to zero at the center of the trap i.e ($E'(r=0)$). Further the expression of $\bar{r}$ at extremum point can be obtained, as given below, 
\begin{equation}
\bar{r}^3=\frac{2b_2(1+c^2)}{a^2}  \label{2d_rbar}
\end{equation}
Substituting the eq.~\ref{2d_rbar} in the second derivative of the potential energy gives the minimum potential depth for the trapping.
\begin{equation}
E''(r)=\frac{a^2}{2(1+c^2)}+\frac{2b_2a^2}{2b_2(1+c^2)}=\frac{3}{2}\frac{a^2}{1+c^2}
\end{equation}
Further expanding potential energy $E(r)$ using Taylor series expansion gives the distribution of potential near the centre of the trap. 
\begin{equation}
E(r)=E(r_0)+\frac{(r-r{_0})^2}{2}E''(r_0)
\end{equation}
Substituting $E''(r_0)$ and $E(r_0)$  in the above equation gives the potential distribution in terms of system parameters, 
\begin{equation}
E(r)=(\frac{1}{4}\frac{a^2\bar{r}^3}{c^2+1}+b)\frac{1}{r_0}+\frac{(r-r{_0})^2}{2}\frac{3}{2}\frac{a^2}{1+c^2} \label{Er}
\end{equation}
It can be observed that the first term of the eq.~\ref{Er} is a constant. Similar to the single droplet motion the probability distribution associated with the given potential energy for two droplet system can be written as,    
\begin{equation}
p(r)=exp[\frac{-E(r)}{K_BT}]=exp[-\frac{a^2(r-r{_0})^2 m \omega^2}{K_BT(1+c^2)}\frac{3}{4}] \label{prob}
\end{equation}
The variance of distribution can be obtained by comparing the expression of probability distribution (eq.~\ref{prob}) with standard form of Gaussian distribution, is given below,  
\begin{equation}
\sigma{_r}^2=\frac{2K_BT(1+c^2)}{3a^2 m \omega^2}   \label{2d}
\end{equation}
Similar to the case of single drop, the eq.~\ref{2d} can be represented in terms of system parameter, 
\begin{equation}
\sigma{_r}^2=\frac{1}{3}[\frac{L_0}{a}]^2 \frac{h}{a} (1+c^2)  \label{sig2d}
\end{equation}
In one of our recent publications \cite{singh2018theoretical} it was found that the presence of inter drop Columbic interactions does not alter the critical stability limit, i.e., the stability of two drops is same as a single drop. Hence it is interesting to check the effect of the presence of noise on the stability of a two droplet system. Numerical simulations show that within the range of physically realizable strength of noise the stability of two droplet does not change. Similar to the analysis of a single drop in a quadrupole trap with fluctuations, the comparison of the mean variance of radial inter drop distance ($\langle\sigma_r^2 \rangle$) for different values of the stability parameter ($a$) with numerical simulation and theoretically obtained expression is given in fig.~\ref{fig:two_n}. 
It can be noticed that the theoretical expression of $\langle\sigma_r^2 \rangle$ for two drop system is 1/3 of a single drop. The fig.~\ref{fig:sig2_vs_a} shows that the nature and stability of the two droplet system is same as single droplet system (fig.~\ref{fig:progeny2}) at a given value of $L_0$. 
In the case of two or more number droplet with a similar charge to mass ratio, the structural symmetry doesn't change in the absence of any external random force. The competition between Columbic interaction force and the time averaged pondermotive force gives an ordered structure onto the XY plane at the equilibrium state. Each drop in a trap oscillates around its equilibrium mean position due to a time varying AC field. A stochastic external force can alter the stability of the equilibrated Columbic structure and gives non-zero variance of inter drop distance. In the absence of noise the structure resembles like a solid whereas in the free moving state when the noise dominates, the structure can transform into a liquid like state. The transition from ordered to disordered state is loosely termed here as the \textquotedblleft phase transition\textquotedblright. A measure of the phase transition or structural disorder is given by the \textquotedblleft Order parameter\textquotedblright. In the case of two droplet levitation, there exist two length scales. First, the change in the average inter drop separation ($\bar{r}$) in the absence of external force which is a unique characteristic of two or a few number of drops (\textless10). Second, the standard deviation ($\sigma$) of the mean inter drop separation in the presence of random external force. Hence, for a larger number of trajectories (\textgreater 1000) the ratio of \textless$\sigma$\textgreater to $\bar{\langle r \rangle}$ is chosen as an order parameter to characterize the phase transition. The numerical simulations of the ensemble average of mean inter drop separation with respect to the strength of fluctuation are shown in fig.~\ref{fig:rbar_vs_Lo}. It can be observed from the figure that initially as the strength of fluctuations is increased up to $L_0$=0.08 the value of $\bar{\langle r \rangle}$ is almost constant and shows a negligible effect of $L_0$. After a particular value of $L_0$ i.e $L_0$$\sim$0.1 the value of $\bar{\langle r \rangle}$  increases at faster rate ($\sim L_0^{0.6}$). Although not shown here it is observed that after $L_0$\textgreater1.3 the droplet motion becomes highly unstable and $\bar{\langle r \rangle}$ increases to a very high value. Thus for $L_0$\textgreater0.1 the $\bar{\langle r \rangle}$ varies with $L_0^{0.6}$ while the eq.~\ref{sig2d} shows that \textless$\sigma^2$\textgreater is proportional to the $L_0^2$. Hence the ratio of \textless$\sigma$\textgreater to $\bar{\langle r \rangle}$ is plotted in the fig.~\ref{fig:sig_r_vs_lo} and the numerically obtained values are compared with analytical theory. It can be observed form fig.~\ref{fig:sig_r_vs_lo} that for $L_0$\textless0.1 the slopes of the curve is $\sim$1 because $\bar{\langle r \rangle}$ is independent of $L_0$ in this region, as shown in fig.~\ref{fig:rbar_vs_Lo}. It is in agreement with  eq.~\ref{sig2d}. In the case of $L_0$\textgreater0.1, $\bar{\langle r \rangle}$ varies with $L_0^{0.6}$, as shown in fig.~\ref{fig:rbar_vs_Lo}, and \textless$\sigma$\textgreater varies with $L_0^{1}$. The resultant slop should be 0.4 and the same is obtained by numerical simulations, as shown in fig.~\ref{fig:sig_r_vs_lo}. The change in the slope at point $L_0=0.1$ (in fig.~\ref{fig:sig_r_vs_lo}) can be considered as the transition from solid like structure to a liquid like structure. The idea of considering the two drop system as a minimal drop system for characterization of phase transition is for the ease of computation. Also, the assumption of symmetry is valid for any number of drops in the case of no fluctuations. A similar idea can be extended to more than two drop system. 
\begin{figure}[htb]
	\centering
	\onefigure[width=0.4\textwidth]{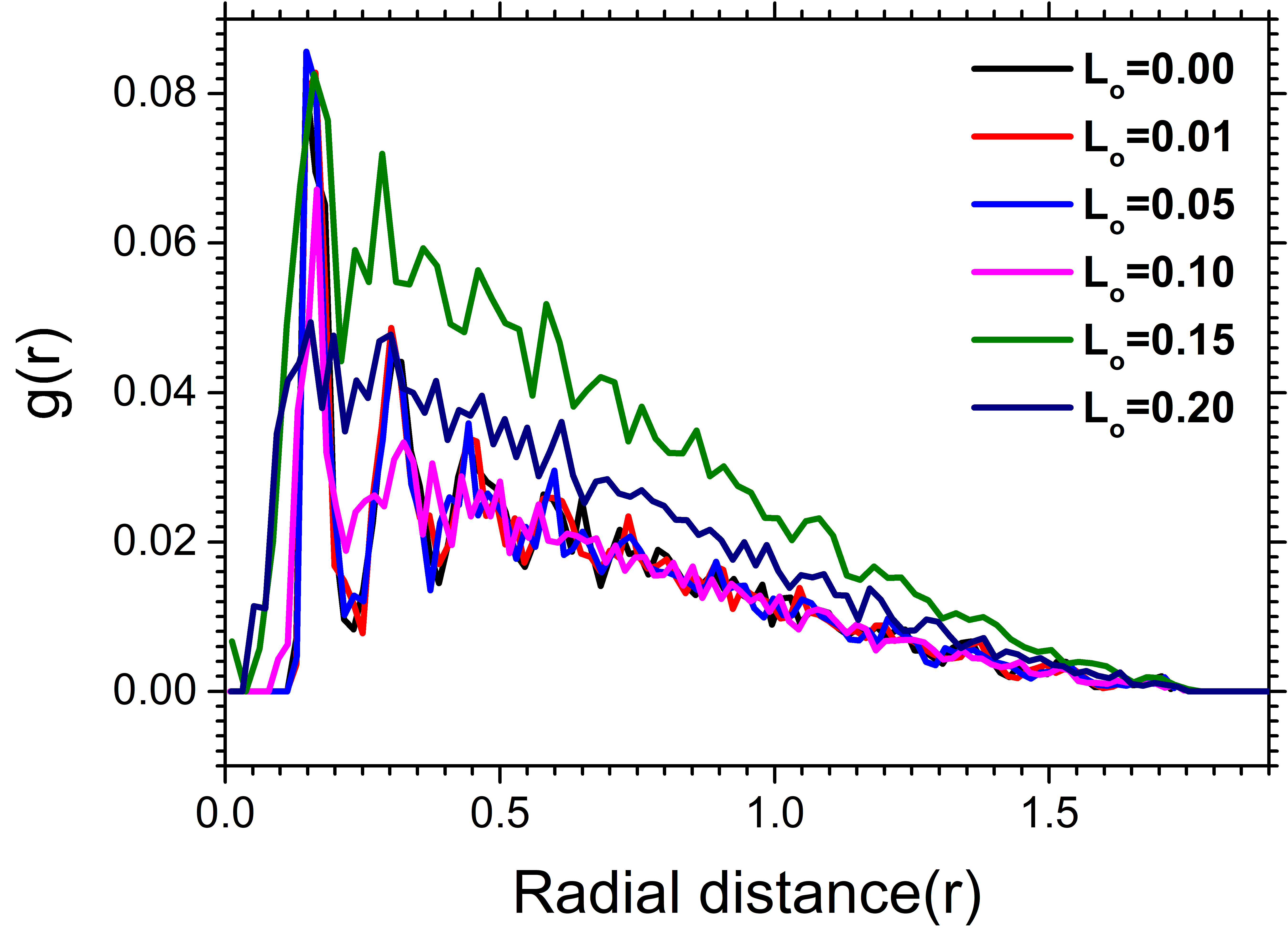}
	\caption{ The radial density distribution of 100 number of drops with varying strength of noise.}
	\label{fig:gofr1}
\end{figure}
The order parameter \textless$\sigma$\textgreater/$\bar{\langle r \rangle}$  is a useful parameter for the characterization of few number of drops (\textless 10). However, for a large number of drops the computational time reaches up to $\sim$ 3-6 days for a single set of parameter on an i7 processor. In case of a large number of drops, there is a lack of a single scalar such as inter droplet distance in the two drop system, to characterize the system. It is therefore pertinent to characterize the system with a different parameter to identify the change in the spatial distribution of the drops. The distribution of drops in the space can be easily characterized by a radial density distribution, $g(r)$. The $g(r)$ distribution can be obtained by calculating the distance between all pairs of drops and binned them into a histogram normalized to the density of the system. Thus the $g(r)$ gives a measure of the spatial correlation of the drops within the system since it is proportional to the probability of finding a drop at a given radial distance from another one. The radial distribution function for 100 number of drops with and without noise is shown in fig.~\ref{fig:gofr1}. 
The spatial positions of all the 100 drops are obtained by taking time average over last 50000 iterations. It can be observed from fig.~\ref{fig:gofr1} that when the value of $L_0$ is $0$ or $0.01$ the $g(r)$ has sharp peaks which can be interpreted as well organized Coulombic cluster. When the value of $L_0$ is further increased the peaks of $g(r)$ start disappearing and a transition from a well organized Coulombic crystal like structure to an amorphous or fluid like structure is observed at a value of $L_0$ \textgreater0.1. A similar transition point is obtained in the case of two drop system ($\sim$ ref fig.~\ref{fig:sig_r_vs_lo}) where a change in the slope is observed at a value of $L_0$\textgreater0.1.             

\section{Conclusions}
The study investigates the effect of noise, induced either by background gas molecular collisions or by electric field fluctuations, on the particles levitated in an electrodynamic balance. Both the case of single and multiple drops are considered, and analytical approximations are derived for the deviations in position from the equilibrium positions. In the case of a single drop, the analytical formula compares well with the present numerical simulations. The results are also compared with the experimental results and numerical simulations of Arnold s. \etal \cite{arnold1995}. The analytical theory has been developed for the effect of noise on the equilibrium positions of self-organized multiple drop structures in an ED balance for the first time, and it is in close agreement with numerical simulations, for values of $a$\textless0.4 at $c$=1 and deviates by 10\% at $a$\textless0.5 at $c$=1. The simulations also indicate the existence of a threshold value of the strength of noise ($L_0=0.1$) above which an organised Coulomb structure undergoes a transition to a random, amorphous structure. 
Our previous experimental observation have shown that the many-body system levitated in an electrodynamic balance has strong stability against various shear forces to which it could be subjected. The experiments have also shown that the many-body system stays stable against the small magnitude of free air pressure drop. Thus it makes the many-body system an appropriate device to use as a steady object for understanding the mechanism by which droplet capture the airborne particles or contaminants flowing through it. Since electrospray technology is being considered as an alternative to corona technology (see ref~\cite{Tepper2006}) for air cleaning understanding the capture of particles by the droplets is very important. The study can be systematically carried in controlled conditions in an ED balance. Alternatively there is also a possibility that the many-body system levitated in an ED balance offers a contactless membrane which can capture the particle directly flowing through the interties under a convectional flow. This requires the system to the stable against perturbations. Thus, the present study has specific relevance to understand the limits of strength of noise which can destabilizes Coulombic structures formed during the levitation in technological application.
\\
%
\acknowledgments
Authors would like to Acknowledge Dr Rakesh Vaiwala and Mr. Inderdip Shere for their help in the code development. Also, we would like to thanks to BRNS, India, for funding the project.    


\end{document}